# Myopic Coding in Wireless Networks

Lawrence Ong and Mehul Motani
Department of Electrical & Computer Engineering
National University of Singapore
E-mail: {lawrence.ong,motani}@nus.edu.sg

*Abstract*— We investigate the achievable rate of data transmission from sources to sinks through a multiple-relay network. We study achievable rates for *omniscient coding*, in which all nodes are considered in the coding design at each node. We find that, when maximizing the achievable rate, not all nodes need to "cooperate" with all other nodes in terms of coding and decoding. This leads us to suggest a constrained network, whereby each node only considers a few neighboring nodes during encoding and decoding. We term this *myopic coding* and calculate achievable rates for myopic coding. We show by examples that, when nodes transmit at low SNR, these rates are close to that achievable by omniscient coding, when the network is unconstrained . This suggests that a myopic view of the network might be as good as a global view. In addition, myopic coding has the practical advantage of being more robust to topology changes. It also mitigates the high computational complexity and large buffer/memory requirements of omniscient coding schemes.

## I. INTRODUCTION

Wireless networks have been receiving much attention recently by both researchers and industry. The main advantage to users of wireless technology is the seamless access to the network whenever and wherever they are. The main advantage to providers of wireless technology is easier deployment as no cable laying is required. These advantages come at the expense of other problems. Data transmission in peer-to-peer wireless networks is done over a shared medium. Hence direct transmission from the source node to a far situated destination node is not desirable as it consumes high transmission power (due to path loss) and creates much interference to other users. Hence data is usually transmitted via multiple-hop routing.

In multiple-hop routing, the common approach in existing works is that the wireless network is abstracted into a communication graph, essentially turning it into a collection of point-to-point links. However, this approach ignores the inherent broadcast nature of the wireless channel, namely that other nodes can hear (and thus can act as relays) transmissions meant for other nodes. To understand how functions such as medium access, routing and transport should be done in "true" wireless networks, we need to understand how to communicate on these wireless networks. This is the aim of this work, to understand how to efficiently communicate data from sources to sinks through a network of wireless relays via cooperation among the nodes.

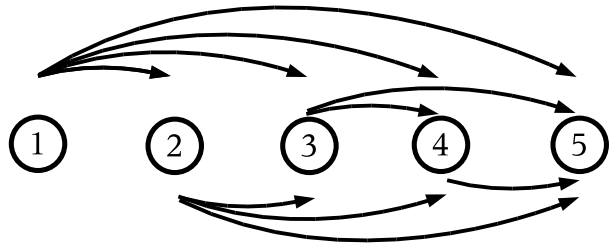

Fig. 1. Omniscient coding on a five-node Gaussian multiple-relay channel.

### A. Coding for the Multiple-Relay Channel

As described, multi-hop routing is desirable in the wireless network. When only one source node and one destination node is being considered, and for a pre-defined fixed route, the scenario reduces to a multiple-relay channel [1]. A five-node multiple-relay channel is depicted in Figure 1, where the leftmost node is the source, the rightmost node is the sink and nodes in the middle relay information for the source.

Clearly, the best thing to do is for all nodes to cooperate to help the source send its data to the sink. This requires a node to be aware of the presence of other nodes and to have knowledge of the processing they do. We call this unconstrained communication on the multiple-relay channel with a global view and complete cooperation *omniscient coding*.

In the literature [2][3], various strategies, including amplify-forward (AF), decode-forward (DF) and compress-forward (CF), for communication on the multiple relay channel are proposed and corresponding achievable rate regions are found. A common characteristic of these strategies is that the coding and decoding at each node takes into account the transmission of all other nodes. Consider a 5-node Gaussian multiple-relay channel, as depicted in Figure 1. Using DF, a node splits its transmission power and sends a fraction of its transmission to each node *in front* of it (towards the destination). For decoding, a node decodes signals from all nodes *behind* (towards the source). At the same time, it cancels interfering transmission from nodes in front. This is possible since it knows what those nodes send, by the direction of information flow. Clearly, the achievable rates for AF, DF and CF are all lower bounds to the best possible rate with omniscient coding.

We discover that in Gaussian multiple-relay channels, using the DF strategy, some of the power splits in omniscient coding are optimum (in the sense of maximizing rate) when set to zero. Details will be given in section III. This suggests that nodes should not transmit to all other nodes. In other words, not all nodes need to "cooperate" with all other nodes to help the source send efficiently to the sink. We call this constrained communication on the multiple-relay channel with a local view and limited cooperation *myopic coding*. To investigate myopic coding, we start by studying achievable rates when a node only transmits to or cooperates with a few neighboring nodes. For example, a two-hop myopic coding scheme is depicted in Figure 2, whereby a node only transmit to two nodes in front.

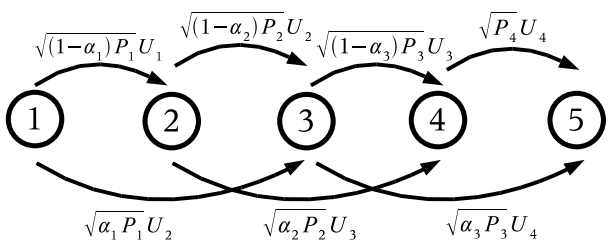

Fig. 2. Two-hop myopic coding on a five-node multiple-relay channel.

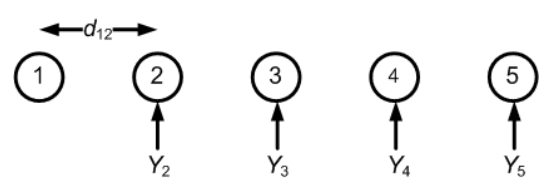

Fig. 3. A five-node Gaussian multiple-relay channel.

### B. Practical Advantages of Myopic Coding

Myopic coding trades performance (though not significantly as we will see later) for some clear practical advantages. In a large network, constructing a coding scheme that takes into account all nodes can be complicated and optimizing the code is more difficult compared to a coding scheme in which a node only transmits to neighboring nodes. This technique of utilizing local knowledge (or limited cooperation) is prevalent in other wireless network problems, e.g., cluster-based routing [4], whereby nodes are split into clusters and routes are optimized locally.

With omniscient coding, any topology change in the network, for example node failure or node mobility, requires reconfiguration of the coding and decoding processes at every node. However, with myopic coding, the failure of one node will only affect its neighboring nodes, thus limiting the reconfiguration required.

Besides being robust to topology changes, myopic coding enjoys several complexity advantages. Since a node only needs to transmit to and decode from a few nodes, there is less computation required in the encoding and decoding processes. Furthermore, since nodes need to buffer data for data transmission and interference cancellation, there is also less memory required for buffering and codebook storage.

### C. Contributions

First, we study omniscient coding in multiple-relay channels in Section III. We consider a five-node Gaussian multiple-relay channel and calculate an achievable rate under omniscient coding. We maximize the achievable rate (when the DF strategy is used) with respect to power splits and show that some power splits can be zero, which means that a node should not transmit to all nodes but only to a few nodes.

Next, we derive achievable rates for one-hop and two-hop myopic coding in multiple-relay channels in Sections IV and V respectively. We show how the rates can be achieved in Shannon-sense via non-constructive coding using the DF strategy. We also extend the achievable rate to that of $k$-hop myopic coding in Section VI.

In Section VII, we compare the achievable rates under omniscient coding and myopic coding. We show that in the five-node Gaussian multiple-relay channel, when the nodes operate in the low transmit signal-to-noise (SNR) region, the achievable rate region under myopic coding is close to that achievable under omniscient coding. This suggests that, in practice, local cooperation is good enough in a large network.

## II. SYSTEM MODEL

In this paper, we investigate omniscient coding and myopic coding on a $T$-node multiple-relay channel, with node 1 being the source node and node $T$ being the destination node. Nodes 2 to $T-1$ are purely relay nodes. Message $W$ is generated at node 1 and is to be transferred to the sink at node $T$. A memoryless multiple-relay channel can be completely described by the channel distribution

$$p^*(y_2, y_3, \dots, y_T | x_1, x_2, \dots, x_{T-1}) \quad (1)$$

on $\mathcal{Y}_2 \times \mathcal{Y}_3 \times \cdots \times \mathcal{Y}_T$, for each $(x_1, x_2, \dots, x_{T-1}) \in \mathcal{X}_1 \times \mathcal{X}_2 \times \cdots \times \mathcal{X}_{T-1}$.

In this paper, we only consider memoryless channels which means

$$p^*(\mathbf{y}_2^n, \mathbf{y}_3^n, \dots, \mathbf{y}_T^n | \mathbf{x}_1^n, \mathbf{x}_2^n, \dots, \mathbf{x}_{T-1}^n)$$
$$= \prod_{i=1}^n p^*(y_{2,i}, y_{3,i}, \dots, y_{T,i} | x_{1,i}, x_{2,i}, \dots, x_{T-1,i}) \quad (2)$$

where $\mathbf{x}_j^n = (x_{j,1}, x_{j,2}, \dots, x_{j,n})$ is an ordered vector of $x_j$ of size $n$.

For comparison, we calculate achievable rates under different coding schemes on a one-dimensional five-node Gaussian multiple-relay channel. The setup is depicted in Figure 3. Node 1 is the source node, nodes 2, 3, 4 are the relay nodes and node 5 is the destination node. Node $i$, $i \in \{1, 2, 3, 4\}$, sends $X_i$ and node $t$, $t \in \{2, 3, 4, 5\}$, receives

$$Y_t = \sum_{\substack{i=1 \\ i \neq t}}^{4} \sqrt{k d_{it}^{-\eta}} X_i + Z_t, \quad (3)$$

where $X_i$ is a random variable with $E[X^2] \leq P_i$, $P_i$ is the power constraint of node $i$, and $Z_t$ is the receiver noise, which is a zero mean Gaussian random variable with variance $N_t$. We assume $X_i$ to be Gaussian. We use a simplified path loss model for signal propagation, in which $\eta$ is the path loss exponent ($\eta \geq 2$ with equality for free space transmission), $k$ is a positive constant, and $d_{it}$ is the distance between node $i$ and node $t$.

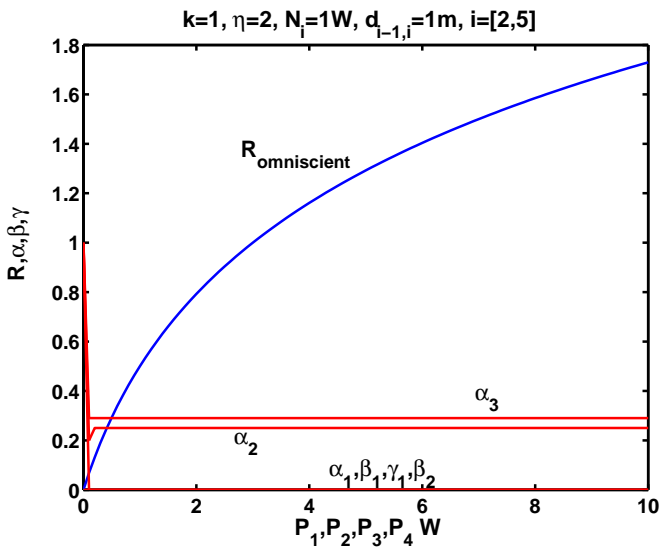

Fig. 4. Achievable rates under omniscient coding in a five-node multiple-relay channel, with equal distance among nodes.

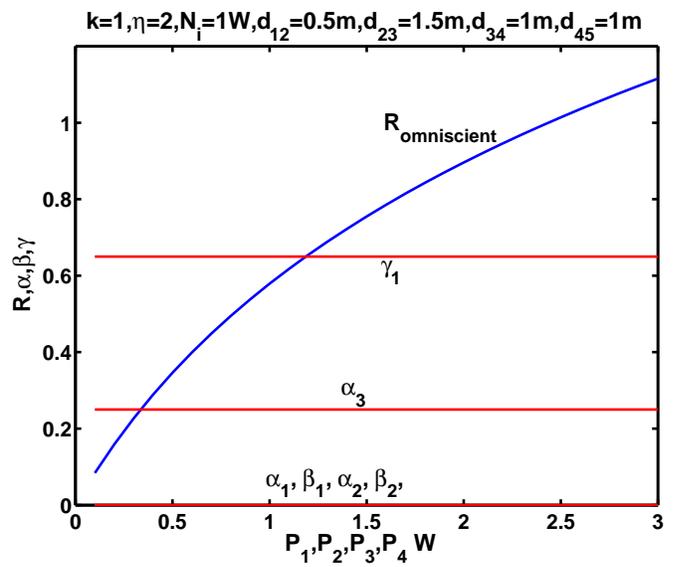

Fig. 5. Achievable rates under omniscient coding in a five-node multiple-relay channel, with node 2 closer to node 1.

## III. ACHIEVABLE RATES WITH OMNISCIENT CODING

Coding based on DF and windowed decoding was proposed by Xie and Kumar [1], whereby a node splits its power and transmits a portion of its power to every node in front. It gets new data from nodes behind itself. Let $\mathcal{R}$ be the set of all relay nodes, $\mathcal{R} = \{2, 3, \ldots, T-1\}$. Let $\pi(\cdot)$ be a permutation on $\mathcal{R}$. Define $\pi(1) = 1$, $\pi(T) = T$ and $\pi(i : j) = \{\pi(i), \pi(i+1), \ldots, \pi(j)\}$. [1] shows that the following rate, which is higher than that in [2], is achievable:

$$R \leq \max_{\pi(\cdot)} \max_{p(\cdot)} \min_{t \in \{1,\ldots,T-1\}} I(X_{\pi(1:t)}; Y_{\pi(t+1)} | X_{\pi(t+1;T-1)}). \quad (4)$$

The outer maximization is over the order of the relay nodes through which data flows. The second maximization is over all possible distributions $p(x_1, x_2, \ldots, x_{T-1})$ on $(X_1, X_\mathcal{R})$. The minimization is on the rate at which each relay node receives.

### A. On Gaussian Channels

On Gaussian channels, the encoding method is as follows:
1) Node 4 sends $X_4 = \sqrt{P_4} U_4$.
2) Node 3 sends $X_3 = \sqrt{(1-\alpha_3)P_3} U_3 + \sqrt{\alpha_3 P_3} U_4$.
3) Node 2 sends $X_2 = \sqrt{(1-\alpha_2-\beta_2)P_2} U_2 + \sqrt{\beta_2 P_2} U_3 + \sqrt{\alpha_2 P_2} U_4$.
4) Node 1 sends $X_1 = \sqrt{(1-\alpha_1-\beta_1-\gamma_1)P_1} U_1 + \sqrt{\gamma_1 P_1} U_2 + \sqrt{\beta_1 P_1} U_3 + \sqrt{\alpha_1 P_1} U_4$.

Here, $U_1, U_2, U_3,$ and $U_4$ are independent Gaussian random variables with unit variances, $0 \leq \alpha_1 + \beta_1 + \gamma_1 \leq 1$, $0 \leq \alpha_2 + \beta_2 \leq 1$, and $0 \leq \alpha_3 \leq 1$. For instance, node 1 allocates $\alpha_1$ of its total power to transmit to node 5, $\beta_1$ of its power to node 4, $\gamma_1$ of its power to node 3, and the remaining power to node 2.

In one-dimensional Gaussian channels, we can show that the achievable rate in (4) is

$$R = \max_{\{\alpha_i, \beta_i, \gamma_i\}} \min_{t \in \{2,\ldots,T\}} R_t, \quad (5)$$

where $R_t$ is the *reception rate* at node $t$ given by

$$R_t \leq \frac{1}{2} \log 2\pi e \left[ k \sum_{j=2}^{t} \left( \sum_{i=1}^{j-1} \sqrt{d_{it}^{-\eta} \alpha_{i,j} P_i} \right)^2 + N_t \right]$$
$$- \frac{1}{2} \log 2\pi e N_t \quad (6a)$$
$$= \frac{1}{2} \log \left[ 1 + \frac{k}{N_t} \sum_{j=2}^{t} \left( \sum_{i=1}^{j-1} \sqrt{d_{it}^{-\eta} \alpha_{i,j} P_i} \right)^2 \right]. \quad (6b)$$

Figures 4 and 5 shows the achievable transmission rates with omniscient coding. Two cases are studied: when all nodes are separated equally and when node 2 is nearer to node 1.

In Figure 4, with equal node spacing, we see that $\alpha_1 = \beta_1 = \gamma_1 = \beta_2 = 0$. This means node 2 only transmits to node 3 and 5; while node 1 only transmits to node 2. In Figure 5, with unequal spacing, $\alpha_1 = \beta_1 = \alpha_2 = \beta_2 = 0$. This means node 2 only transmits to node 3; while node 1 only transmits to nodes 2 and 3. This selective transmission suggests that a node should not cooperate with all nodes in its coding and decoding.

The fact that a node need not cooperate with all other nodes to maximize the achievable rate leads us to investigate achievable rates when nodes can only transmit to a few other nodes. A systematic way is to investigate a constrained network where each node can only "see" a few neighboring nodes, i.e., nodes are myopic, choosing only to interact with nodes close to themselves. The motivation here is that if the achievable rates of the constrained network are as good as the unconstrained network, the simpler and more practical approach of myopic coding may be good in large networks.

## IV. ONE-HOP MYOPIC CODING

In one-hop myopic coding, each node only sends signals to the node in front of it and decode signals from the node behind it. With one-hop coding, node $t$ can receive information up to rate

$$R_t \leq \max I(X_{t-1}; Y_t | X_t) \quad (7)$$

for $t \in \{2, \ldots, T\}$ and $X_T = 0$. The maximization is over the distribution $p(x_1)p(x_2) \cdots p(x_{T-1})$. Since all information must pass through all nodes in order to reach the destination, the overall rate is constrained by

$$R = \min_{t \in \{2,\ldots,T\}} R_t. \quad (8)$$

## V. TWO-HOP MYOPIC CODING

Instead of just transmitting to one node in front, a node might want to help the node in front to transmit to the node that is two hops away. We term this two-hop myopic coding, where a node transmits to nodes within two hops away. We consider $B + T - 2$ transmission blocks, each of $n$ uses of the channel. A sequence of $B$ independent indices, $w^b \in \{1, 2, \ldots, 2^{nR}\}$, $b = 1, 2, \ldots, B$ will be sent over $n(B + T - 2)$ uses of the channel. As $B \to \infty$, the rate $RnB/n(B + T - 2) \to R$ for any $n$.

### A. Codebook Generation and Encoding

In this section, we see how codebooks at each node are generated.

1) First, fix the probability distribution

$$p(u_1, u_2, \ldots, u_{T-1}, x_1, x_2, \ldots, x_{T-1})$$
$$= p(u_1)p(u_2) \cdots p(u_{T-1})p(x_1|u_1, u_2)p(x_2|u_2, u_3)$$
$$\cdots p(x_{T-1}|u_{T-1})$$

for each $u_i \in \mathcal{U}_i$.
2) For each $t \in \{1, \ldots, T-1\}$, generate $2^{nR}$ independent and identically distributed (i.i.d.) $n$-sequences in $\mathcal{U}_t^n$, each drawn according to $p(\mathbf{u}_t) = \prod_{i=1}^n p(u_{t,i})$. Index them as $\mathbf{u}_t(w_t)$, $w_t \in \{1, \ldots, 2^{nR}\}$.
3) Define $\mathbf{x}_{T-1}(w_{T-1}) = \mathbf{u}_{T-1}(w_{T-1})$.
4) For each $t \in \{1, \ldots, T-2\}$, define a deterministic function that maps $(\mathbf{u}_t, \mathbf{u}_{t+1})$ to $\mathbf{x}_t$:

$$\mathbf{x}_t(w_t, w_{t+1}) = f_t(\mathbf{u}_t(w_t), \mathbf{u}_{t+1}(w_{t+1})). \quad (10)$$

Here, the subscript $t$ for $w_t$ indicate the new message transmitted by node $t$.
5) In block $b \in \{1, \ldots, B + T - 2\}$, assuming node $t$, $t \in \{1, \ldots, T-1\}$, has decoded $(w^1, w^2, \ldots, w^{b-t+1})$, sends $\mathbf{x}_t(w^{b-t+1}, w^{b-t})$. Here, we use superscript to indicate the time index of the source letter, meaning that the source emits $w^1, w^2, \ldots, w^B$ at the beginning of block $1, 2, \ldots, B$ respectively. The encoding for the first few block of nodes 1 to 4 is depicted in Figure 7.

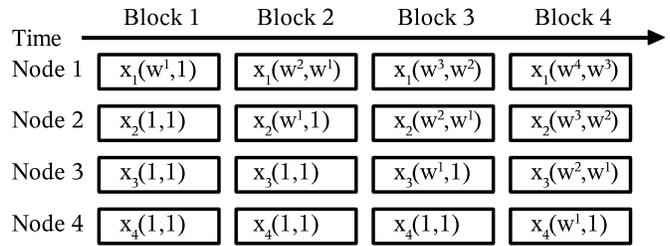

Fig. 7. A two-hop encoding scheme

We see that in each transmission block, node $t$, $t \in \{1, \ldots, T-2\}$, sends two message indices $w_t$ (new data) and $w_{t+1}$ (old data). In the same block, node $t+1$ sends messages $w_{t+1}$ and $w_{t+2}$. Note that node $t$ cooperates with the node $t+1$ by repeating the transmission $w_{t+1}$.

### B. Decoding

The decoding of a source letter is carried out over two blocks. Referring to Figure 6, in block $b-1$, node $t-2$ sends $\mathbf{x}_{t-2}(w^{b-t+2}, w^{b-t+1})$, node $t-1$ sends $\mathbf{x}_{t-1}(w^{b-t+1}, w^{b-t})$. Knowing $w^{b-t+1}$, $w^{b-t}$, and $w^{b-t-1}$, node $t$ find a set of $w^{b-t+2}$ for which

$$\{\mathbf{u}_{t-2}(w^{b-t+2}), \mathbf{u}_{t-1}(w^{b-t+1}), \mathbf{u}_t(w^{b-t}), \mathbf{u}_{t+1}(w^{b-t-1}), \mathbf{y}_t\}$$
$$\in \mathcal{A}_\epsilon^n(U_{t-2}, U_{t-1}, U_t, U_{t+1}, Y_t). \quad (11)$$

Here, $\mathcal{A}_\epsilon^n(S)$ represents the set of $\epsilon$-typical $n$-sequence of the random variables in the set $S$. We follow the definition of $\epsilon$-typical $n$-sequences defined in [5]. In block $b$, node $t-1$ sends $\mathbf{u}_{t-1}(w^{b-t+2}, w^{b-t+1})$. Knowing $w^{b-t+1}$ and $w^{b-t}$, node $t$ find a set of $w^{b-t+2}$ for which

$$\{\mathbf{u}_{t-1}(w^{b-t+2}), \mathbf{u}_t(w^{b-t+1}), \mathbf{u}_{t+1}(w^{b-t}), \mathbf{y}_t\}$$
$$\in \mathcal{A}_\epsilon^n(U_{t-1}, U_t, U_{t+1}, Y_t). \quad (12)$$

Node $t$ then finds the intersection of the two sets to determine the value of $w^{b-t+2}$. This can be done reliably if $R \leq I(U_{t-2}, U_{t-1}; Y_t | U_t, U_{t+1})$. This is only the rate constraint at one node. In two-hop coding using the DF strategy, each message must be fully decoded at each node. Hence the overall rate is constrained by

$$R = \min_{t \in \{2,\ldots,T\}} R_t \quad (13)$$

where

$$R_t \leq I(U_{t-2}, U_{t-1}; Y_t | U_t, U_{t+1}) \quad (14)$$

is the reception rate at node $t$. We fix $U_0 = U_T = U_{T+1} = 0$. This gives us the following theorem.

*Theorem 1:* Consider a $T$-node multiple-relay channel with transition probability

$$p^*(y_2, \ldots, y_T | x_1, \ldots, x_{T-1}).$$

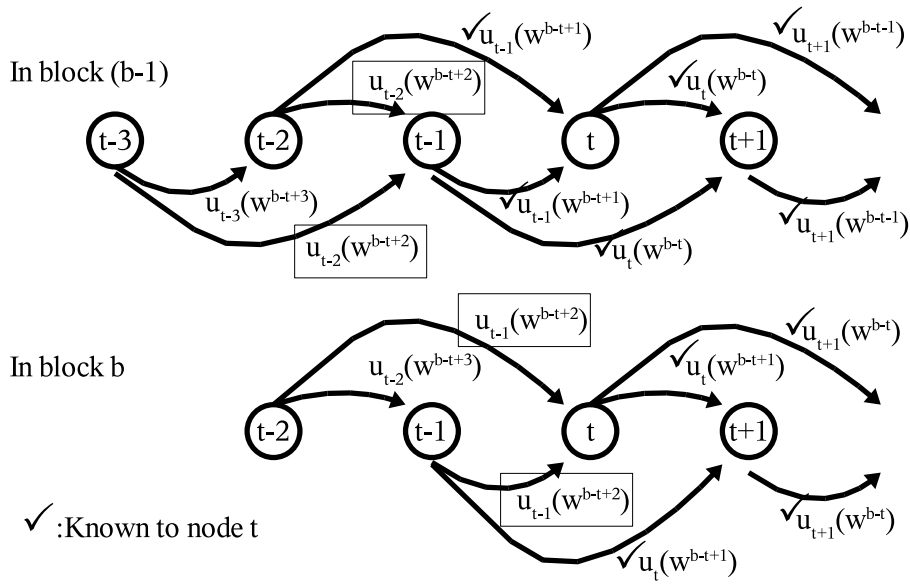

Fig. 6. Decoding at node $t$ of message $w^{b-t+2}$

Under two-hop coding where each node only transmits to two nodes in front and decode a message over two blocks, the rate $R$ is achievable, where

$$R \leq \sup \min_{t\in\{2,...,T\}} I(U_{t-2}, U_{t-1}; Y_t | U_t, U_{t+1}) \quad (15)$$

where $U_0 = U_T = U_{T+1} = 0$ and the supremum is taken over all joint distribution of the form

$$p(x_1, x_2 \ldots, x_{T-1}, u_1, u_2 \ldots, u_{T-1}, y_2, y_3 \ldots, y_T)$$
$$= p(u_1)p(u_2) \cdots p(u_{T-1})p(x_1|u_1, u_2)p(x_2|u_2, u_3) \cdots$$
$$p(x_{T-1}|u_{T-1})p^*(y_2, \ldots, y_T | x_1, \ldots, x_{T-1}).$$

## VI. $k$-HOP MYOPIC CODING

We define *$k$-hop myopic coding* as a constrained communication in the multiple-relay channel where a node can only transmit to $k$ neighboring nodes and decode a message symbol over $k$ blocks. We can show that the following rate is achievable under $k$-hop myopic coding using the DF strategy.

*Theorem 2:* Consider a $T$-node memoryless multiple-relay channel with channel with channel transition probability

$$p^*(y_2, \ldots, y_T | x_1, \ldots, x_{T-1}).$$

Under $k$-hop coding where each node only transmits to $k$ nodes in front, the rate $R$ is achievable, where

$$R \leq \sup \min_{t\in\{2,...,T\}} I(U_{t-k}, \ldots, U_{t-1}; Y_t | U_t, \ldots, U_{t+k-1}) \quad (17)$$

where $U_{2-k} = U_{3-k} = \cdots = U_0 = U_T = U_{T+1} = \cdots = U_{T+k-1} = 0$ and the supremum is taken over all joint distribution of the form

$$p(x_1, x_2 \ldots, x_{T-1}, u_1, u_2 \ldots, u_{T-1}, y_2, y_3 \ldots, y_T)$$
$$= p(u_1)p(u_2) \cdots$$
$$\times p(u_{T-1})p(x_{T-1}|u_{T-1})p(x_{T-2}|u_{T-2}, u_{T-1}) \cdots$$
$$\times p(x_{T-k}|u_{T-k}, u_{T-k+1} \ldots, u_{T-1})$$
$$\times p(x_{T-k-1}|u_{T-k-1}, u_{T-k} \ldots, u_{T-2}) \cdots$$
$$\times p(x_1|u_1, u_2, \ldots, u_k)$$
$$\times p^*(y_2, \ldots, y_T | x_1, \ldots, x_{T-1}).$$

## VII. COMPARISON ON GAUSSIAN CHANNELS

In this section, we compare achievable rates for one-hop myopic coding, two-hop myopic coding, and omniscient coding on Gaussian multiple-relay channels. On Gaussian channels, with two-hop myopic coding, node $t, t = 1, 2, 3$, allocate $\alpha_t$ of its power to transmit to node $t+2$ and $(1-\alpha_t)$ of its power to node $t+1$. Since there is only one node in front of node 4, it transmits only to node 5. The transmission by each node is listed as follows:

1) Node 4 sends $X_4 = \sqrt{P_4}U_4$.
2) Node 3 sends $X_3 = \sqrt{\alpha_3 P_3}U_4 + \sqrt{(1-\alpha_3)P_3}U_3$.
3) Node 2 sends $X_2 = \sqrt{\alpha_2 P_2}U_3 + \sqrt{(1-\alpha_2)P_2}U_2$.
4) Node 1 sends $X_1 = \sqrt{\alpha_1 P_1}U_2 + \sqrt{(1-\alpha_1)P_1}U_1$.

Here, $U_i, i = 1, 2, 3, 4$ are independent Gaussian random variables with unit variances and $0 \leq \alpha_1, \alpha_2, \alpha_3 \leq 1$.

Figures 8 and 9 show the achievable rate under one-hop coding, two-hop coding and omniscient coding. Two node configurations are studied, that is when all nodes are separated equally and when node 2 is nearer to node 1.

When the nodes are equally spaced, the achievable rate with omniscient coding is always larger than than that achievable with two-hop coding. This is intuitive because in myopic coding, interactions among the nodes are constrained and this might restrict the achievable rate. However, when node 2 is closer to node 1, the achievable rate with two-hop coding is close, or even equal (in low SNR region), to that achievable under omniscient coding.

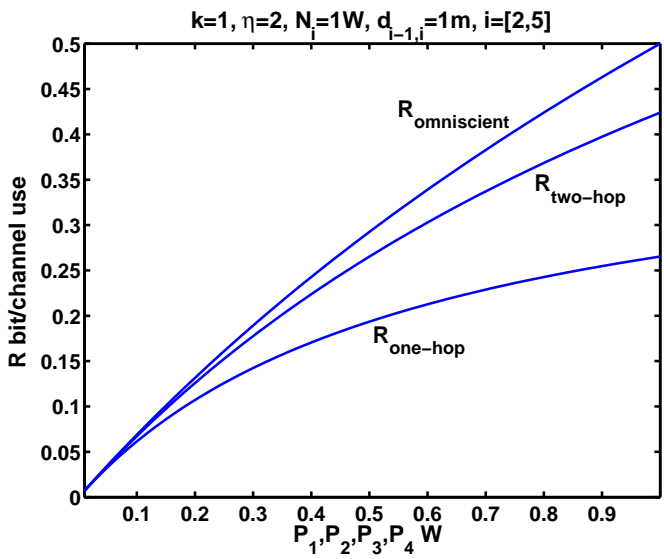

Fig. 8. Achievable rates comparison under the different schemes in a five-node multiple-relay channel, with equal distance among nodes.

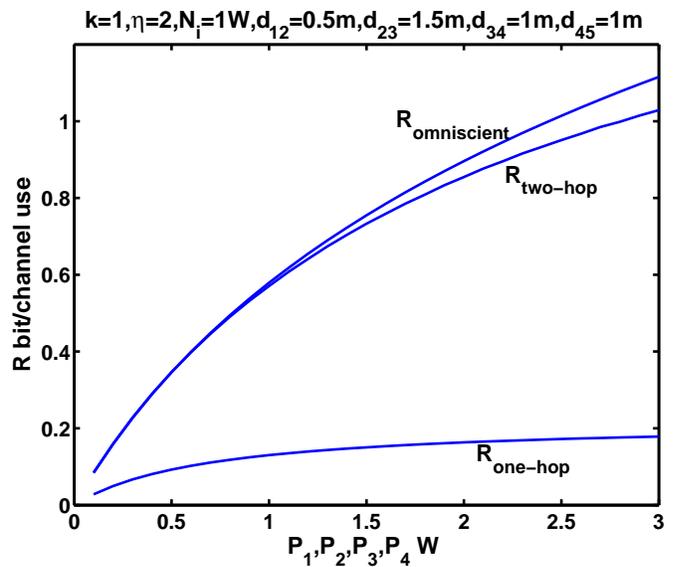

Fig. 9. Achievable rates comparison under different schemes in a five-node multiple-relay channel, with node 2 closer to node 1

It is noted the achievable rate region with two-hop coding is as large as that with omniscient coding only when the overall transmission rates in both cases are constrained by the reception rate at the same node and when the number of the nodes in the channel is small, such that the node can cancel all interference even with myopic view.

In Figure 9, the achievable rate with one-hop coding is low. This is because as $d_{23}$ is set to 1.5 m, the reception rate at node 3 is penalized and it constrains the overall achievable rate. By adding just another node to the view (increasing from one-hop coding to two-hop coding), we see an significant increase in the achievable rate. Also, we should expect diminishing returns as more nodes are added into the view as transmission between two far away nodes are attenuated due to path-loss. Our results suggest that coding with local view is good enough in large networks.

## VIII. Conclusion

We have found an achievable rate region for myopic coding on the multiple-relay channel, where cooperation among the nodes is constrained. We have shown that in a five-node Gaussian multiple-relay channel, when nodes transmit at low SNR, the achievable rate region with two-hop myopic coding is almost as large as that achievable under omniscient coding. We also see a significant increase in achievable rates when comparing one-hop myopic coding and two-hop myopic coding, meaning that we might not need to increase a node's view much farther than a few nodes. Hence, besides having practical advantages, myopic coding is potentially (as only non-constructive coding is considered in this paper) as good or close to omniscient coding. This means in a large network, we could possibly limit the cooperation and perform local coding design without compromising much on the transmission rate.

The analysis in this paper helps us to understand communication and cooperation in the multiple-relay channel better. This work sheds light on how one might design practical and efficient transmission protocols in wireless networks, where robustness, computational power, and storage memory are important design considerations, in addition to transmission rate.

## References

[1] L. Xie and P.R. Kumar, "An achievable rate for the multiple level relay channel," *IEEE Trans. Inform. Theory.*, submitted in Nov. 2003.
[2] P. Gupta and P.R. Kumar, "Towards an information theory of large network: an achievable rate region," *IEEE Trans. Inform. Theory.*, vol. 49, no. 8, pp. 1877–1894, Aug. 2003.
[3] G. Kramer, M. Gastpar and P. Gupta, "Cooperative strategies and capacity theorems for relay networks", *IEEE Trans. Inform. Theory.*, submitted in Feb. 2004.
[4] Y. T. Mingliang Jiang and Jinyang Li, "Cluster based routing protocol," IETF Internet Draft, draft-ietf-manet-cbrp-spec-01.txt, July 1999. [Online]. Available: http://www.comp.nus.edu.sg/∼tayyc/cbrp/.
[5] T.M. Cover and J.A. Thomas, "Elements of Information Theory," John Wiley and Sons, 1991.